\documentclass[10pt,a4paper]{book}
\usepackage{udineHyderabad-Klu2}

\begin{document}

	\pagestyle{fancy}
	
\talktitle{Grid services for the MAGIC experiment}{Grid services for the MAGIC experiment}

\shorttitle{Grid services for the MAGIC telescope}

\talkauthors{A.~Forti\structure{1}, S.R.~Bavikadi\structure{1}, C.~Bigongiari\structure{2}, G.~Cabras\structure{1}, A.~De~  Angelis\structure{1}, B.~De~Lotto\structure{1}, M.~Frailis\structure{1}, M.~Hardt\structure{3}, H.~Kornmayer\structure{3}, M.~Kunze\structure{3}, M.~Piraccini\structure{1} and the MAGIC Collaboration}
  
\authorstucture[1]{INFN and Dipartimento di Fisica, Universit\'a degli
  Studi di Udine, Italy}
  
\authorstucture[2]{INFN and Dipartimento di Fisica Galileo Galilei, Padova, Italy}

\authorstucture[3]{Institut f\"{u}r wissenschaftliches Rechnen, Forschungszentrum Karlsruhe, Germany}

\firstauthor{A.~Forti et al.}

		\index{Forti@\textsc{Forti}, A.}
		\index{Bavikadi@\textsc{Bavikadi}, S.R.}
		\index{Bigongiari@\textsc{Bigongiari}, C.}
		\index{Cabras@\textsc{Cabras}, G.}
		\index{De Angelis@\textsc{De Angelis}, A.}
		\index{De Lotto@\textsc{De Lotto}, B.}
		\index{Frailis@\textsc{Frailis}, M.}
		\index{Hardt@\textsc{Hardt}, M.}
		\index{Kornmayer@\textsc{Kornmayer}, H.}
		\index{Kunze@\textsc{Kunze}, M.}
		\index{Piraccini@\textsc{Piraccini}, M.}
		\index{MAGIC Collaboration@\textsc{MAGIC Collaboration}}

\begin{abstract}
  Exploring signals from the outer space has become an observational
  science under fast expansion. On the basis of its advanced
  technology the MAGIC telescope is the natural building block for the
  first large scale ground based high energy $\gamma$-ray observatory.
  The low energy threshold for $\gamma$-rays together with different
  background sources leads to a considerable amount of data. The
  analysis will be done in different institutes spread over Europe.
  Therefore MAGIC offers the opportunity to use the Grid technology to
  setup a distributed computational and data intensive analysis system
  with the nowadays available technology. Benefits of Grid computing
  for the MAGIC telescope are presented.
\end{abstract}

\section{MAGIC}

The MAGIC telescope has been designed to search the sky to discover or
observe high energy $\gamma$-rays sources and address a large number
of physics questions \protect\cite{ref1}. Located at the Instituto
Astrophysico de Canarias on the island La Palma, Spain, at 28° N and
18° W, at altitude 2300m asl, it is the largest $\gamma$-ray telescope
in the world. MAGIC is operating since October 2003, data are taken
regularly since February 2004 and signals from Crab and Markarian 421
was seen.
The main characteristics of the telescope are summarized below:
\begin{itemize}
\item A 17m diameter (f/d=1) tessellated mirror mounted on an
  extremely light carbon-fiber frame ($<10$ tons), with active mirror
  control. The reflecting surface of mirrors is 240 $m^2$;
  reflectivity is larger than 85\% (300 - 650nm).
\item Elaborate computer-driven control mechanism.
\item Fast slewing capability (the telescope moves 180° in both axes
  in 22s).
\item A high-efficiency, high-resolution camera composed by an array
  of 577 fast photomultipliers (PMTs), with a 3.9° field of view.
\item Digitalization of the analogue signals performed by 300 MHz
  FlashADCs and a high data acquisition rate of up to 1 KHz.
\item MAGIC is the lowest threshold ($\approx30$ GeV) IACT operating
  in the world.
\end{itemize}
Gamma-rays observation in the energy range from a few tenth of GeV
upwards, in overlap with satellite observations and with substantial
improvement in sensitivity, energy and angular resolution, leads to
search behind the physics that has been predicted and new avenues will
open. AGNs, GRBs, SNRs, Pulsars, diffuse photon background,
unidentified EGRET sources, particle physics, darkmatter, quantum
gravity and cosmological $\gamma$-ray horizon are some of the physics
goals that can be addressed with the MAGIC telescope.

\section{Grid}

The idea of computational and data grids dates back to the first half
of the 90's. The vision behind them is often explained using the
electric power grid metaphor. The electric power grid delivers
electric power in a pervasive and standardized way. You can use any
device that requires standard voltage and has a standard plug if you
are able to connect it to the electric power grid through a standard
socket.  When you use electricity you don't worry were it is produced
and how it is delivered, you just plug your device into the wall
socket and use it. Currently we have millions of computing and storage
systems all over the planet connected through the Internet. What we
need is an infrastructure and standard interfaces capable of providing
transparent access to all this computing power and storage space in a
uniform way. This is the concept behind Grid. More precisely, Grid is
a kind of parallel and distributed system that enables the sharing,
selection and aggregation of services of heterogeneous resources
distributed across multiple administrative domains based on their
availability, capability, performance, cost, and users
quality-of-service requirements \protect\cite{ref2}. As network performance
has outpaced computational power and storage capacity, this new
paradigm has evolved to enable the sharing and coordinated use of
geographically distributed resources.

\subsection{Virtual Organization}

The Virtual organization is an important Grid concept. Grid allows a
pool of heterogeneous resources both within and outside of an
organization to be virtualized and form a large, virtual computer.
This virtual computer can be used by a collection of users and/or
organizations in collaboration to solve their problems. The rules
governing the participants providing the resources and the consumers
using the resources, as well as the conditions for sharing, dictate
the nature of the virtual organization. Hence, a virtual organization
groups people and resources without worry about their physical
location or institute boundaries. For security reasons, the use of the
resources is constrained by an authentication process.

\section{Benefits of Grid computing for Magic}

The collaborators of the MAGIC telescope are mainly spread over
Europe, 18 institutions from 9 countries, with the main contributors
(90\% of the total) located in Germany (Max-Planck-Institute for
Physics, Munich and University of Wuerzburg), Spain (Barcelona and
Madrid), Italy (INFN and Universities of Pa\-do\-va, Udine and Siena).

The geographical distribution of the resources makes the management of
the experiment harder. This is a typical situation for which Grid
computing can be of great help, because it allows researchers to
access all the resources in a uniform, transparent and easy way.  The
telescope is in operation during moonless nights. The average amount
of raw FADC data recorded is about 500-600 GB/night. Additional data
from the telescope control system or information from a weather
station are also recorded. All these information have to be taken into
account in the data analysis.

The MAGIC community can leverage from Grid facilities in areas like
file sharing, Monte Carlo data production and analysis. In a Grid
scenario the system can be accessed through a web browser based
interface with single sign-on authentication method. We can briefly
summarize the main benefits given by the adoption of Grid technology
for the MAGIC experiment:
\begin{itemize}
\item Presently, users analyzing data must know where to find the
  required files and explicitly download them. In a Grid perspective,
  instead, users don't care about data location and files replication
  policies improve access time and fault tolerance.
\item Grid workflow tools can manage the MAGIC Monte Carlo simulation.
  The resources from all the members of the MAGIC community can be put
  together and exploited by the Grid. Easy access to data production
  for every user, or accordingly to the VO policies.
\item Analysis tools can be installed on the Grid. They are thus
  shared and available for all the users (no need for single
  installations), moreover, they can exploit the facilities of a
  distributed system.
\end{itemize}

\subsection{MMCS}
The MAGIC Monte Carlo simulation workflow is a series of programs
\begin{figure}
  \begin{center}
    \includegraphics[scale=0.33]{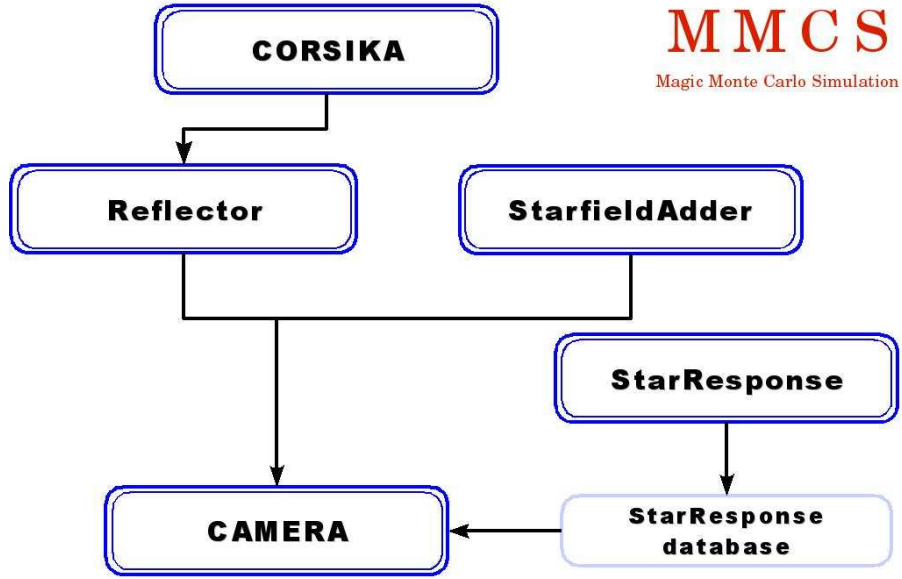}%
    \caption{MAGIC Monte Carlo Simulation workflow}
    \label{MMCS}
  \end{center}
\end{figure}
which simulate the properties of different physics processes and
detector parts (figure \ref{MMCS}):
\begin{itemize}
\item \emph{CORSIKA}: air shower and hadronic background simulation.
  The output contains information about the particles and the
  Cherenkov photons reaching the ground around the telescope.
\item \emph{Reflector}: simulates the propagation of Cherenkov photons
  through the atmosphere and their reflection in the mirror up to the
  camera plane. The input for the Reflector program is the output of
  CORSIKA.
\item \emph{StarfieldAdder}: simulation of the field of view. It adds
  light from the non-diffuse part of the night sky background, or the
  effect of light from stars, to images taken by the telescope.
\item \emph{StarResponse}: simulation of NSB (night sky background)
  response.
\item \emph{Camera}: simulate the behavior of the photomultipliers and
  of the electronic of the MAGIC camera. It also allows to introduce
  the NSB (optionally), from the stars and/or the diffuse NSB.
\end{itemize}

\subsection{The present}

The EGEE (Enabling Grids for E-science in Europe) project brings
together experts from 70 organizations and 27 countries with the
common aim of building on recent advances in Grid technology and
developing a service grid infrastructure in Europe which is available
to scientists 24 hours-a-day. Recently MAGIC has became part of the
EGEE project. This is the first step for enabling MAGIC on the Grid.
This process migration is a big effort and must be divided in smaller
steps.  The first step was chosen to be the migration of the MAGIC
Monte Carlo simulation workflow and it is now working \protect\cite{ref3}
thanks to the effort of the Udine, Padova and CNAF (Bologna) groups.

\section{Conclusions}

Grid technologies promise to change the way organizations tackle their
complex computational and data-intensive problems. The vision of large
scale resource sharing is nowadays becoming a reality in many areas.
However, it must be realized that Grid is an evolving field in
computer science, where standards and technology are still being under
development to enable this new paradigm. Presently, many efforts are
being made to attract a wide range of new users to the grid. MAGIC has
caught this big opportunity and now another step is made towards the
realization of a wide project that wants to connect the compute and
storage resources of the astroparticle institutions in order to
collaborate across the institute borders as well as across the
collaboration borders. This new challenge will be the ASTROPA Grid
project \protect\cite{ref4}.

\end{document}